\documentclass[10pt]{rcdj}

\def\prooffont{\sc}
\rcdthmsep=\medskipamount

\renewrcdthm{teo}{Теорема}
\renewrcdthm{lem}{Лемма}
\renewrcdthm{cor}{Следствие}
\renewrcdthm{pro}{Предложение}
\renewrcdthm[definition]{fed}{Определение}
\renewrcdthm[remark]{rem}{Замечание}
\renewrcdthm[definition]{com}{Комментарий}
\renewrcdthm[example]{exl}{Пример}

\renewcommand{\thesection}{\arabic{section}}

\begin{document}
\journal{REGULAR AND CHAOTIC DYNAMICS, V.\,8, \No2, 2003}

\setcounter{page}{163}

\received 11.11.2002.

\amsmsc{37J35, 70H06, 76B47}

\doi{10.1070/RD2003v008n02ABEH000235}

\title{AN INTEGRABILITY OF THE PROBLEM ON MOTION OF CYLINDER AND VORTEX IN
THE IDEAL FLUID}

\runningtitle{AN INTEGRABILITY OF THE PROBLEM ON MOTION OF CYLINDER AND VORTEX}
\runningauthor{A.\,V.\,BORISOV, I.\,S.\,MAMAEV}

\authors{A.\,V.\,BORISOV, I.\,S.\,MAMAEV}
{Institute of Computer Science\\
Universitetskaya, 1\\
426034, Izhevsk, Russia\\
E-mail: borisov@rcd.ru\\
E-mail: mamaev@rcd.ru}
\abstract{In this paper we present the nonlinear Poisson structure and two first
integrals in the problem on plane motion of circular cylinder and $N$
point vortices in the ideal fluid. A priori this problem is not
Hamiltonian. The particular case ${N=1}$, i.\,e. the problem on
interaction of cylinder and vortex, is integrable.
}
\maketitle

\section{The equations of motion}

Let's consider the problem on plane motion of cylindrical rigid body
and~$n$ point vortices with~circulations~$\Gamma_i$ in an~unbounded volume
of the ideal incompressible fluid motionless in infinity. We assume that
the exterior force fields are absent, the surface of cylinder is ideally
smooth, and~the cylinder flow is circulating, i.\,e. the circulation along
a closed contour encircling the cylinder is ${\Gamma\ne0}$. The equations
of motion of such system were almost simultaneously obtained by
S.\,M.\,Ramodanov in~\cite{ram} and~also in~\cite{shashi}. In~\cite{ram} it
was assumed that $n=1 $ and~in~\cite{shashi} that $\Gamma=0$.
Paper~\cite{ramo} is the expanded version of~\cite{ram} presenting the
most general equations of motion of body and~vortices ($\Gamma\ne 0$ and
$n$ is arbitrary). In~the latter paper it is supposed that~$\Gamma=0$.
In~the further text we adhere to paper \mbox{\cite{ramo}}.

Let's consider two frames of references: the motionless frame~$OXY$
and~the one connected with~the center of the cylinder~$Cxy$ performing a
plane-parallel motion with respect to the first frame
(Fig.~\ref{vort-01.eps}). Let $(v_1,v_2)= \bv$ be the projections of
velocity of the center of cylinder onto the axes of~$Cxy$,
$(x_i,y_i)=\br_i$ the projections of radius-vector connecting the center
of cylinder with~$i$\1th vortex onto the same axes, $\mu$ the mass of
cylinder, $R$ its radius. Paper~\cite{ram} shows that for the case of the
inertial motion, equations for~$\bv$,~$\br_i$ are separated from the whole
system and~can be integrated independently. They have the following form
\eqc[ur1]{
\left.\dot{\br}_i=-\bv{+}\grad\wt\vfi_i\right|_{\br=\br_i},\\
\mu\dot v_1=\lm v_2{-}\suml_{i=1}^n\lm_i(\dot{\wt y}_i{-}\dot y_i),\;
\mu\dot v_2={-}\lm v_1{+}\suml_{i=1}^n\lm_i(\dot{\wt x}_i{-}\dot x_i),
}
where $\wt\br_i=(\wt x_i,\wt y_i)$ is the inverse image of the $i$\1th
vortex conjugated with respect to the contour of cylinder by the rule $\wt
x_i=\frac{R^2}{\br^2}\,x_i$, $\wt y_i=\frac{R^2}{\br^2}\,y_i$ and
function~$\wt\vfi_i$ corresponds to the part of flow potential~$\vfi$ that
has no singularities at~$\br=\br_i$. The flow potential outside the
cylinder can be presented in the form
\begin{equation*}
\vfi(\br)=-\frac{R^2}{\br^2}\,(\br,\bv)-\lm\arctg\frac yx
+\suml_{i=1}^n\lm_i\Bigl(\arctg\Bigl(\frac{y-\wt y_i}{x-\wt x_i}\Bigr)-
\arctg\Bigl(\frac{y-y_i}{x-x_i}\Bigr)\Bigr),
\end{equation*}
where $\lm=\frac\Gamma{2\pi}$, $\lm_i=\frac{\Gamma_i}{2\pi}$.

\fig<bb=0 0 62.9mm 55.2mm>{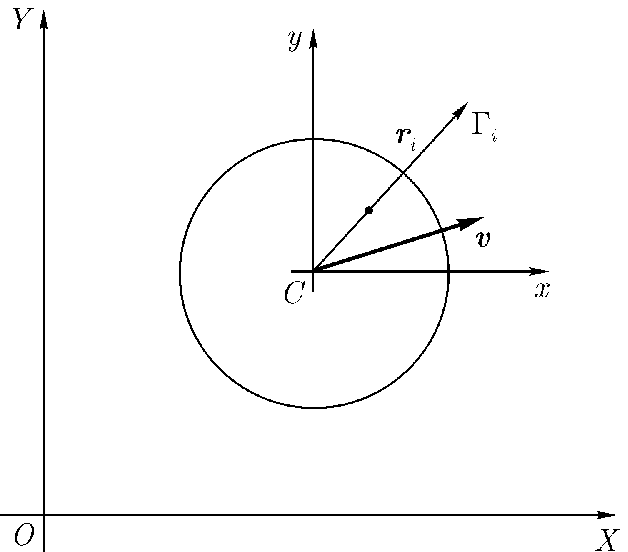}

It is easy to verify that the equations~\eqref{ur1} have an integral
(similar to the energy integral if we use an analogy with classical
mechanics) in the form
\begin{equation}
\label{ur2}
H=\frac12\,\mu(v_1^2+v_2^2)+\frac12\suml_i
[\lm_i^2\ln(r_i^2-R^2)-\lm_i\lm\ln r_i^2]
+\frac12\suml_{i<j}\lm_i\lm_j\ln
\frac{R^4-2R^2(r_i,r_j)+r_i^2r_j^2}{|r_i-r_j|^2},
\end{equation}
though the origin of this integral not obvious since~\eqref{ur1} are not
derived as usual from the Lagrangian formalism, but~inherit some
conservative properties from both the dynamics of rigid body and~the
motion of the ideal fluid. Note that~\eqref{ur1} preserve the standard
invariant measure.

\section{A Hamiltonian structure of the equations of motion}

The problem of existence of Hamiltonian form for~\eqref{ur1} was
considered in paper~\cite{shashi} and some results were obtained for the
case~${\Gamma=0}$, $\suml_{i=1}^n\Gamma_i=0$. Here we present the Poisson
structure for~\eqref{ur1} in the general case. It is nonlinear with
respect to the phase variables and nondegenerate. Its basis nonzero
Poisson brackets are the following
\begin{alignat}{2}
\{v_1,v_2\}&{=}\frac\lm{\mu^2}{-}
\sum\frac{\lm_i}{\mu^2}\frac{r_i^4-R^4}{r_i^4},&\;
\{v_1,x_i\}&{=}\frac1\mu\frac{r_i^4{-}R^2(x_i^2{-}y_i^2)}{r_i^4},\notag\\
\label{ur3}
\{v_1,y_i\}&=-\frac1\mu\frac{2R^2x_iy_i}{r_i^4},&\;
\{v_2,x_i\}&=-\frac1\mu\frac{2R^2x_iy_i}{r_i^4},\\
\{v_2,y_i\}&=\frac1\mu\frac{r_i^4+R^2(x_i^2-y_i^2)}{r_i^4},&\;
\{x_i,y_i\}&=-\frac1{\lm_i}.\notag
\end{alignat}
According to the Darboux theorem in~the nondegenerate case we can always
find the analytical transformation presenting~\eqref{ur3} in the canonical
form $\{p_i,q_i\}=\dl_{ij}$, $\{p_i,p_j\}= 0$, $\{q_i,q_j\}=0$. However,
the explicit form of this transformation is not required for our analysis.

\section{An integrability of the system for the case of one vortex}

Let's consider the case $n=1$, i.\,e. the simultaneous motion of vortex
and cylinder. Denote $\br_1=\br=(x, y)$. We can easily see that such
system is invariant with respect to the rotation about the center of
cylinder. Therefore, there is the integral
\eq[ur4]{
F=\mu'\bv^2{+}2\mu\lm_1\Bigl(1{-}\frac{R^2}{\br^2}\Bigr)(xv_2{-}yv_1){+}
\lm_1\Bigl(\lm_1\Bigl(\br^2{+}\frac{R^4}{\br^2}\Bigr){-}\lm\br^2\Bigr),{}
}
generating because of~\eqref{ur3} the field of symmetries
\eq[ur5]{
\bv_F=2\lm(v_2-v_1,y_1-x).
}
The integral~\eqref{ur4} is a generalization of the integral of moment
from classical mechanics and using it we can integrate the system in
quadratures. We can reduce the order of system down to one degree of
freedom. This reduction is close to the Routh reduction. Let's perform it
explicitly.

As the variables of the reduced system we shall choose the integrals of the
field of symmetries~$\bv_F$~\eqref{ur5} (see, for example,~\cite{bormam}).
For example,
\eqa[ur6]{
p_1&=\mu(xv_1+yv_2),\q& p_2&=\mu(xv_2-yv_1),\\
r_1&=\mu^2(v_1^2+v_2^2),\q& r_2&=x^2+y^2.
}
Poisson brackets between these variables are the following
\begin{equation}
\label{ur7}
\begin{gathered}
\{p_1,p_2\}=(\lm-\lm_1)r_2+\frac1{\lm_1}(p_1^2+(p_2-\lm_1R^2)^2),\\
\{r_1,r_2\}=4p_1\Bigl(1+\frac{R^2}{r^2}\Bigr),\\
\{p_1,r_1\}=2(\lm-\lm_1)p_2-2\frac{p_1^2+p_2^2}{r_2}+
2R\frac{p_1^2-p_2(p_2-\lm_1R^2)}{r_2^2},\\
\{p_1,r_2\}=2r_2+\frac2{\lm_1}(p_2-\lm_1R^2),\\
\{p_2,r_1\}=-2(\lm-\lm_1)p_1+2R^2p_1\frac{2p_2^2-\lm_1R^2}{r_2^2},\\
\{p_2,r_2\}=-2\frac{p_1}{\lm_1}.
\end{gathered}
\end{equation}
The Poisson structure~\eqref{ur7} is degenerated, its rank is equal to two
and therefore such reduced system has one degree of freedom. To obtain it
in explicit form we shall exclude two variables from the
set~$(r_1,r_2,p_1,p_2)$ using the integral~\eqref{ur4}, the Casimir
function of structure~\eqref{ur7}, and the obvious geometrical relation
${p_1^2+p_2^2-r_1r_2=0}$ following from~\eqref{ur6}.

The qualitative analysis of the reduced system would be an interesting
study that could lead to some conclusions about the motion of the complete
system~\eqref{ur1}.

\section{The case of $n$\1vortices}

In the general case there is a generalization of integral~\eqref{ur4}. It
reads
\begin{equation}
\label{ur8*}
F=\mu\bv^2+\suml_{i=1}^n\lm_i\biggl(2\mu\biggl(1{-}\frac{R^2}{\br_i^2}\biggr)
(x_iv_2{-}y_iv_1){+}(\lm_i{-}\lm)\br_i^2{+}\lm_i\frac{R^4}{\br_i^2}\biggr){+}
2\suml_{i<j}\lm_i\lm_j(\br_i,\br_j)\biggl(1{-}\frac{R^2}{\br_i^2}\biggr)
\biggl(1-\frac{R^2}{\br_j^2}\biggr).
\end{equation}
Using this generalization we can also perform the reduction of the order
of system to one degree of freedom. For the integrability of the obtained
system we shall have $n-1$~involute integrals. In the general case,
probably, they do not exist and the system of two vortexes interacting
with the cylinder is already nonintegrable. However, it is not proved yet
and there may exist the particular cases of integrability under additional
restrictions on parameters of the system and on constants of the known
integrals~\eqref{ur2},~\eqref{ur8*}.

There is another interesting problem on a generalization of Poisson
structure~\eqref{ur2} and integrals~\eqref{ur2}, \eqref{ur8*} for the
problem on interaction in fluid of two (or several) rigid bodies with
the given circulations (certainly, in the plain formulation). Recently the
equations of two cylinders in fluid (without circulations) were obtained
by S.\,M.\,Ramodanov. In such formulation they are ``plane'' analog of the
known Bjerkness problem on interaction of two balls in a flow of ideal
fluid. These results for the first time were published in the collection.
This particular problem is integrable in contrast to the
more general situation when the flow of cylinders is circulating. The
corresponding equations of motion are not obtained yet.

Let's consider also another related problem, which is also integrable. It
is the interaction of elliptic Kirchhoff vortex with point vortex in an
approximation described by the second order theory of moments~(see, for
example,~\cite{bormam}). In contrast to the rigid body, the Kirchhoff
vortex during the motion remains elliptic, but changes the lengths of
semiaxes. The problems on interaction of two Kirchhoff vortexes or one
Kirchhoff vortex with two point vortexes seem to be nonintegrable.

We thank V.\,V.\,Kozlov and S.\,M.\,Ramodanov for useful remarks.

\end{document}